# A 2-bit Wideband 5G mm-Wave RIS with Low Side Lobe Levels and no Quantization Lobe

Ruiqi Wang, Yiming Yang, *Student Member, IEEE*, and Atif Shamim, *Fellow, IEEE*

*Abstract*—Reconfigurable intelligent surface (RIS) with 1-bit phase resolution suffers from high side lobe levels (SLLs) in the near field and pronounced quantization lobe levels (QLLs) in the far field, which is detrimental for the quality of wireless communication. RIS design is further challenging in the mm-wave bands, where typically large bandwidths are required. To address these limitations, this work proposes a novel wideband 5G mm-Wave RIS with 2-bit phase quantization, capable of covering the entire 5G n258 band. The proposed unit cell design is a combination of a grounded main slot, a parasitic slot, and a coupling patch, utilizing only two PIN diodes. This design achieves a 2-bit bandwidth (BW) of 24.1–27.7 GHz (13.9%) and an effective 1-bit BW of 20.0–28.9 GHz (36.4%). The unit cell features a compact size of $0.39\lambda \times 0.39\lambda$, providing decent angular stability of $\pm30°$ as well as eliminating the grating lobes. Based on this unit cell design, a $20 \times 20$ RIS array has been designed, fabricated and experimentally characterized. The measured results demonstrate that, within the 5G n258 band, the proposed 2-bit mm-Wave RIS achieves an SLL of -15.4 dB in the near field, representing a 7.6 dB improvement compared to its 1-bit counterpart. Furthermore, it has almost negligible QLL (-14.6 dB) in the far field, providing a suppression of 13.3 dB relative to the 1-bit design. Thus, the proposed 2-bit mm-Wave RIS offers wideband performance, low SLL, negligible QLL, decent angular stability, and a broad beam scanning range of 50°, making it a promising solution for high-resolution and low-interference mm-Wave wireless communication systems.

*Index Terms*— 2-bit phase quantization, fifth generation (5G), millimeter-wave (mm-Wave), quantization lobe level (QLL), reconfigurable intelligent surface (RIS), side lobe level (SLL), wideband.

## I. INTRODUCTION

FIFTH-GENERATION 5G millimeter-wave (mm-Wave) communication system has been extensively studied due to its large available bandwidth and the potential for substantial increases in communication capacity [1], [2], [3]. However, 5G mm-Wave wireless communication faces significant challenges, such as high path loss and atmospheric attenuation, particularly when the transceivers (Tx) and receivers (Rx) are blocked by obstacles [4]. Recently, reconfigurable intelligent surfaces (RIS) concept has been developed and extensively investigated as a promising solution to these challenges. RIS can establish an alternative line of sight (LOS) to maintain the communication link [5], [6], [7]. By integrating RIS, the electromagnetic (EM) environment becomes digitally programmable, leading to significant improvements in the performance of wireless communication systems [8]. A conceptual diagram of multi-bit 5G mm-Wave RIS applications is shown in Fig. 1, illustrating a scenario that the RIS can solve the signal blockage issue. Moreover, the desired reflection toward User 1 is maximized for high signal enhancement, while interference to other users is minimized.

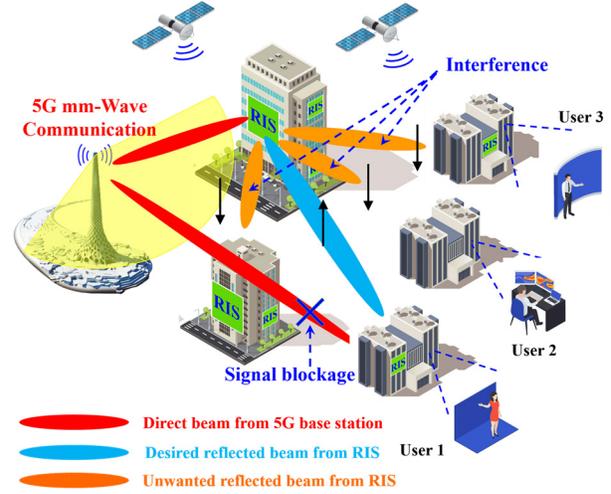

Fig. 1. Conceptual framework of multi-bit 5G mm-Wave RIS applications in wireless communication systems with low interference.

In the literature, extensive studies have been conducted on RIS [9 – 12], highlighting the potential of RIS in enhancing the quality of wireless communication. However, most published works on RISs remain theoretical, lacking experimental validation [9 – 12]. For experimental demonstrations of RIS prototypes, majority of the reported designs operate at sub-6 GHz bands [13 – 22], or frequencies below or around 10 GHz [23 – 25]. There are only a handful of experimental works in the 5G mm-Wave bands [26 – 31]. References [26] and [27] report mm-Wave RIS designs, but their bandwidths (BWs) are narrow, typically less than 7%, which is not sufficient to meet the large BW requirements of 5G communication systems. Other studies have demonstrated mm-Wave RISs for applications such as path loss modeling [28], amplification [29], and beamforming characterization [30], yet they do not

Authors are with the Computer, Electrical and Mathematical Sciences and Engineering Division, King Abdullah University of Science and Technology (KAUST), Thuwal 23955, Saudi Arabia (e-mail: ruiqi.wang1@kaust.edu.sa; yiming.yang@kaust.edu.sa; atif.shamim@kaust.edu.sa). *(Corresponding author: Ruiqi Wang.)*

provide details on the RIS bandwidth. In our previous work, we proposed a wideband 5G mm-Wave RIS that covered the entire 5G n257 and n258 bandwidth [31]. However, it suffers from high side lobe levels (SLLs) in the near field. Moreover, unlike reconfigurable reflectarray antennas [32 – 38], RIS is expected to operate not only in the near field but also in the far field, where significant quantization lobe levels (QLLs) exist for 1-bit RIS [31]. These negative effects of 1-bit phase discretization inadvertently strengthen the channel of the eavesdropper, compromising security in practical wireless networks [39]. Therefore, a multi-bit 5G mm-Wave RIS is highly desirable to mitigate these vulnerabilities effectively.

Typically, multi-bit RIS designs utilize varactors, as a single varactor in each unit cell can achieve nearly 360° of continuous phase shift [21], [25]. However, at 5G mm-Wave frequencies, varactors suffer from low Q factor, significant parasitic capacitance and inductance, higher losses (up to 10 dB [40]). In contrast, PIN diodes offer a viable solution due to their fast switching speeds (~2 ns) and high operational frequencies [41], [42]. However, in multi-bit RIS designs, the cost of PIN diodes becomes a critical factor, as the price of the switches is considerably higher than the cost of printed circuit board (PCB) fabrication. In terms of performance of multi-bit designs, it has been shown that the gain degradation for 1-bit, 2-bit, and 3-bit phase quantization is approximately 3.0, 0.6, and 0.2 dB, respectively, compared to continuous phase distribution [43 – 45]. Thus, a 2-bit 5G mm-Wave RIS design offers a reasonable trade-off between fabrication cost and RIS performance. An additional challenge arises when determining the number of switches needed for 2-bit phase quantization. Some studies have used multiple switches to implement 2-bit phase quantization [5], [46], [47], which significantly increases the fabrication cost for 5G mm-Wave systems. To the best of the authors' knowledge, no existing wideband 2-bit 5G mm-Wave RIS (PIN-diode-based) can cover the n258 frequency band (24.25 – 27.5 GHz, 12.6%) using only two PIN switches due to the difficulties in controlling and tuning multiple resonances.

In this work, we demonstrate a novel 2-bit 5G mm-Wave RIS that utilizes only two PIN switches in each unit cell, achieving a wide 2-bit phase resolution bandwidth (BW) of 13.9%, which fully covers the 5G n258 band. Additionally, it offers an effective 1-bit BW of 36.4%. Compared to the current state-of-the-art 5G mm-Wave RIS [31], the proposed design demonstrates significant improvements, including a 7.6 dB side lobe level (SLL) reduction, 3.7 dB higher signal enhancement at the desired RIS beam reflection in the near field, and a 13.3 dB suppression of quantization lobe level (QLL) in the far field. The 1-bit RIS BW has also increased from 26.6% to 36.4%. Therefore, the proposed RIS design presents a promising solution for 5G mm-Wave wireless communication systems, particularly for applications requiring advanced functionality and low noise.

## II. RIS UNIT CELL DESIGN AND SIMULATION

### A. Design Concept and Evolution

The design is based on multiple slots coupled with a patch,

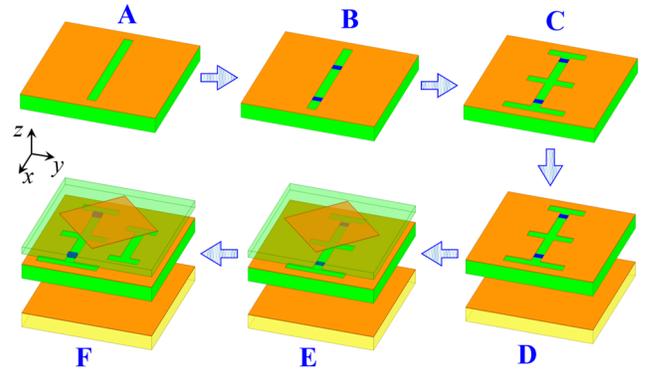

Fig. 2. Evolution of the designed RIS unit cell.

as shown in Fig. 2 with the design evolution. The design starts from a slot structure (Fig. 2A). Two RF switches are placed on the slot aperture to generate four different resonant frequencies (Fig. 2B). The location of the switches should be asymmetric to have four different reconfigurable states. In the next step, three perpendicular cutouts are included in the slot (Fig. 2C) to obtain a larger phase variation range as well as more flexible tuning parameters for these four states. It is worth mentioning that the y-axis cutouts should be symmetric about x-axis to avoid cross-polarization components. The ground is at the bottom of this unit cell (Fig. 2D) as a reflector, and a coupling patch is added on the top of the unit cell to enhance the bandwidth (Fig. 2E). Finally, a parasitic slot is placed beside the main slot (Fig. 2F), which generates another resonant frequency to further enlarge the BW. Notice that the effects of feeding network and the parasitic components of the PIN switches are not included at this stage of the study but will be presented later.

After the design evolution, a wideband 2-bit phase quantization can be achieved using only two switches. To quantitatively illustrate the functions of the parasitic slot and coupling patch, comparisons between the designed RIS unit cells with and without the parasitic slot and coupling patch, are shown in Fig. 3. A wideband phase transition between these four states is observed for the designed unit cell, as demonstrated in Fig. 3(a). Specifically, the phase differences between each state are maintained within $90° \pm 30°$ from 25.0 to 29.5 GHz (16.5% BW). In contrast, when the designed unit cell does not include the parasitic slot, the operational bandwidth becomes narrower, as shown in Fig. 3(b). This is because the parasitic slot introduces a parasitic resonance around 30 GHz, which expands the overall phase transition curve between various configurations. Additionally, without the top coupling patch, the slot resonator exhibits a much narrower bandwidth, as illustrated in Fig. 3(c). For the proposed unit cell design, three y-axis slot segments are extended on the main slot, providing better design and optimization flexibility. The parasitic slot can operate independently, enabling flexible adjustment of the parasitic resonant mode, which provides advantages over self-resonant high harmonic resonances with fixed $2.5f$ and $3.5f$ modes [31].

To further explain the operating principle of the designed unit cell, the surface current distributions at different states are shown in Fig. 4. It can be observed that two resonances at 19.1 GHz and 30.1 GHz in the OFF/OFF state of the two PIN switches (red curve) in Fig. 3(a) are excited by the main slot and parasitic slot, respectively, as illustrated in Fig. 4(a). By configuring the switch states, the electrical length of the main slot is gradually reduced, leading to the main resonance shifting to higher frequencies. Meanwhile, due to the coupling effect between the main slot and the parasitic slot, the parasitic mode also shifts to a higher resonant frequency, as shown in the OFF/ON, ON/OFF, and ON/ON states in Fig. 4(b) and Fig. 3(a). Therefore, a wideband 2-bit phase transition between these four states can be achieved using only 2 PIN switches. The positions of the two PIN switches should be appropriately adjusted to control different resonant modes among these four different reconfigurable states and prevent the generation of unwanted resonances from various segments.

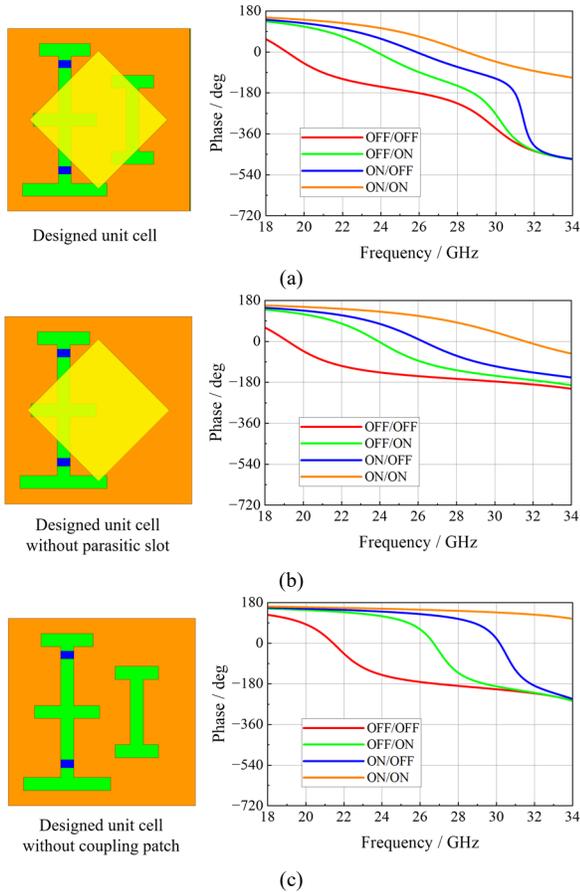

Fig. 3. Simulated $S_{11}$ phase performance of unit cells when switches at various states. (a) Proposed unit cell, (b) without parasitic slot, (c) without coupling patch.

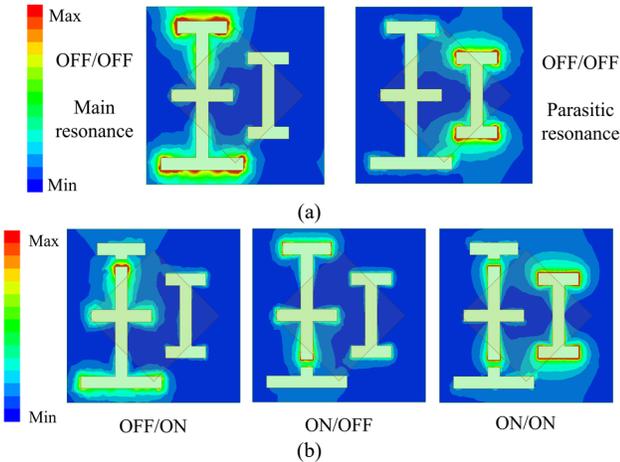

Fig. 4. Surface current distribution of the designed unit cell. (a) Main resonance and parasitic resonance in the OFF/OFF state. (b) Main resonances in OFF/ON, ON/OFF, and ON/ON states.

### B. Practical Design

The developed 2-bit unit cell model, illustrated in Fig. 5(a), is applicable and scalable for lower operational frequencies (<10 GHz). However, at 5G mm-Wave frequencies, the effects of the biasing circuit, feeding vias, networks, and PIN diodes need to be incorporated in the EM simulation. Additionally, practical fabrication feasibility should be considered. Building upon the proposed unit cell model, a practical 2-bit 5G mm-Wave unit cell is introduced in Fig. 5(b), with its development detailed below.

The first step in refining the unit model from Fig. 5(a) is to determine the biasing scheme for the realistic PIN diodes in order to isolate the DC connections within the overall structure. Due to the presence of the y-axis slot segments, the main slot can be separated with simple DC isolation structures, where RF capacitors are placed at the borders of the RF and DC slots to ensure the RF performance remains unaffected by the DC structures, as shown in Fig. 5(b).The entire slot layer serves as the DC common ground connected to the cathode of the PIN diodes, while two DC-isolated pads, connected to the anodes of the PIN diodes, can be independently controlled. The PIN diodes used in this work are the commercially available MA4AGFCP910 type from MACOM, which offers decent performance at mm-Wave frequencies.

The second step involves routing the DC anode pad to connect with the biasing lines, typically through vias. Due to the limited routing space available on the small unit cell size at 5G mm-Wave frequencies, the biasing lines cannot be fully arranged on the backside of the slot layer. Therefore, the slot layer and ground layer are merged into a single stacked board, allowing vias to pass through the ground layer while placing the biasing layer behind it (Fig. 5 (b)) on two sides of a Rogers 5880 laminate with a thickness of 1.575 mm, while the biasing layer is etched on a 0.1 mm-thick FR4 laminate. These two substrates are sandwiched by a 0.102 mm RO4450F prepreg sheet.

The third step involves the placement of the parasitic patch. The distance between the coupling patch substrate and the slot layer in Fig. 5(a) is 0.4 mm ($h_2$), which does not provide sufficient space for soldering the RF capacitors and PIN diodes. Therefore, the coupling patch is relocated from the top side to the bottom side of the substrate, maintaining the proper EM coupling between the slot layer and the coupling patch while providing an adequate 0.8 mm soldering space for the RF components. Moreover, the top 0.508 mm-thick Rogers 5880 substrate serves not only as the RF substrate but also as protection to prevent exposure of the soldered RF components. As a result, the overall RF structures are enclosed within the designed unit cell, making it suitable for future RIS-in-package concepts. Following these refinements, the unit cell dimensions have been finally optimized for the 5G n258 frequency band, as detailed in Table I.

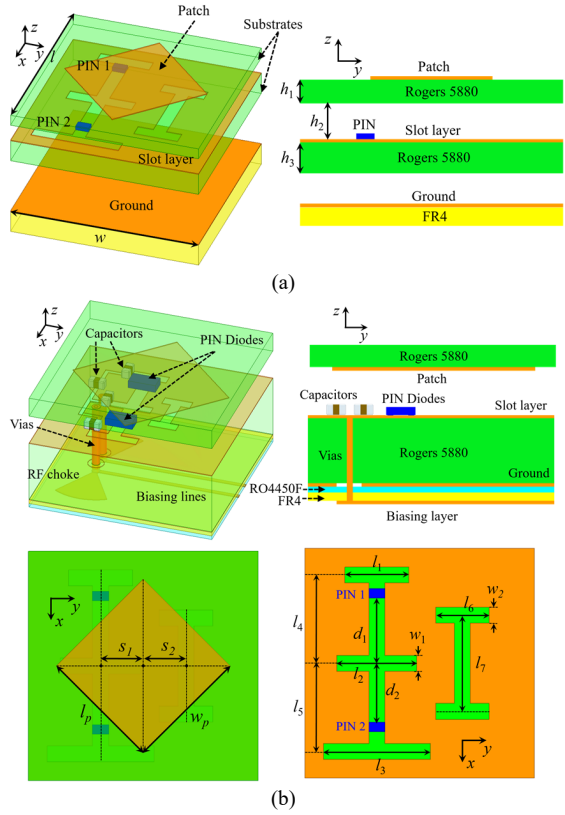

Fig. 5. Proposed 2-bit 5G mm-Wave RIS unit cell design. (a) Designed unit cell model. (b) Practical unit cell design.

TABLE I
DIMENSIONS OF THE DESIGNED PRACTICAL 5G MM-WAVE UNIT CELL (mm)

| $l$ | $w$ | $l_p$ | $w_p$ | $l_1$ | $l_2$ | $l_3$ | $l_4$ | $l_5$ | $l_6$ |
|---|---|---|---|---|---|---|---|---|---|
| 4.6 | 4.6 | 2.78 | 2.78 | 0.88 | 1.0 | 1.25 | 1.42 | 1.42 | 0.7 |
| $l_7$ | $w_1$ | $w_2$ | $d_1$ | $d_2$ | $h_1$ | $h_2$ | $h_3$ | $s_1$ | $s_2$ |
| 1.05 | 0.26 | 0.26 | 0.87 | 0.69 | 0.508 | 0.8 | 1.575 | 0.5 | 0.5 |

### C. Unit Cell Performance

The simulated reflection magnitude and phase performance of the designed practical 5G mm-Wave RIS unit cell at four different configurations are presented in Fig. 6(a) and (b), respectively. The reflection magnitudes and phase transition curves for all four states demonstrate satisfactory performance. To quantitatively assess the 2-bit RIS unit cell performance, the average reflection magnitude and equivalent phase bits are calculated based on the simulation results, as shown in Fig. 7. Specifically, the reflection magnitude of the 2-bit RIS unit cell is defined as the average of the four reconfigurable states, while the equivalent bits are determined using the evaluation criteria outlined in [48], [45]:

$$N_{bit} = \frac{1}{2}\log_2\left(\frac{360^3}{\sum_{i=1}^{4}(\Delta\phi_i)^3}\right) \quad (1)$$

where $\Delta\varphi_i$ denotes the phase difference between two adjacent phase states. Due to practical design imperfections, achieving a strict 90° phase difference between each state over a wide frequency band is very challenging. Therefore, a 2-bit phase quantization is considered effective when the equivalent bits exceed 1.7 [48], which can be used to define the 2-bit frequency bandwidth. For the proposed unit cell design, the average unit cell magnitude remains above -0.76 dB within the frequency range of $24 - 28$ GHz. Additionally, the equivalent bit exceeds 1.7 from 24.1 GHz to 27.8 GHz, providing an operating bandwidth of 14.3%, which fully covers the 5G n258 frequency band.

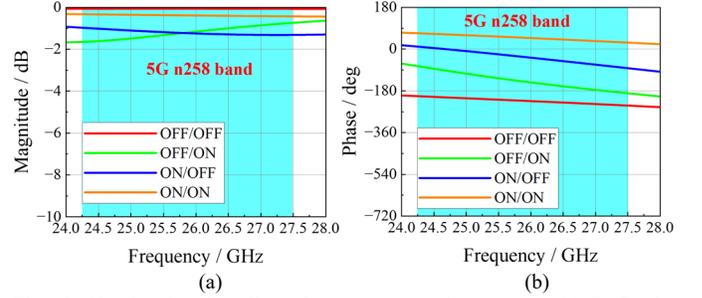

Fig. 6. Simulated unit cell performance at various states. (a) Reflection Magnitude. (b) Reflection phase.

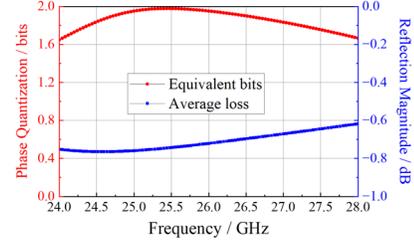

Fig. 7. Average reflection magnitude and the equivalent bits of the designed practical unit cell.

Lastly, for RIS applications, angular stability of the unit cell design is a critical criterion, as the incident signal impinging on the RIS is typically unknown. The reflection magnitude and phase performance of the designed unit cell at various incident angles, ranging from a normal incidence of 0° to an oblique angle of 30° with a step size of 10°, are shown in Fig. 8. From the simulation results, it is evident that both the magnitude and phase responses of the design maintain consistent performance across random incident angles within the 0° to 30° range. This demonstrates the robust properties of the designed unit cell, making it suitable for practical applications.

## II. RIS DESIGN AND SIMULATION

### A. RIS Array Design

The 3D stack-up array configuration of the proposed 2-bit 5G mm-Wave RIS design is illustrated in Fig. 9. The coupling patches are etched on the bottom surface of a 0.508mm-thick Rogers 5880 substrate, while the other board consists of a lamination of Rogers 5880 and FR4 substrates, laminated through an RO4450F prepreg. The lamination board features RF capacitors and switches soldered on the top surface, with connectors located on the bottom surface. The feeding network is routed effectively in a single-layer structure, and the overall RIS design is free of any blind vias. Finally, the coupling patch board and the lamination board are secured together with mechanical supports using screws along the edges. The complete RIS, arranged in a $20 \times 20$ configuration, contains a total of 400 unit cells and has dimensions of $8.0\,\lambda \times 8.0\,\lambda$, where $\lambda$ corresponds to the center operating frequency of 26 GHz. The biasing control for the entire RIS is managed by a specifically designed control circuit.

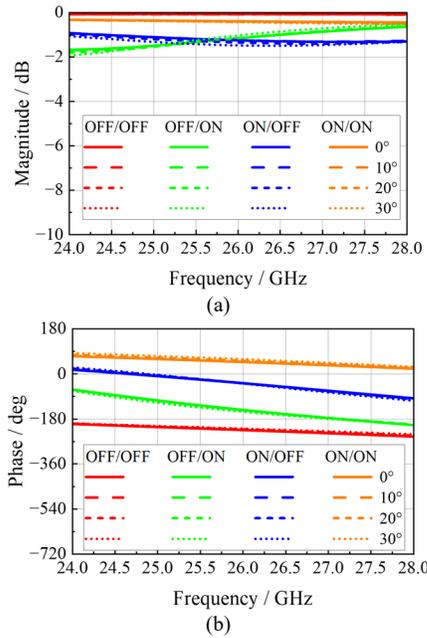

(a)

(b)

Fig. 8. Angular stability of the designed unit cell. (a) Magnitude. (b) Phase.

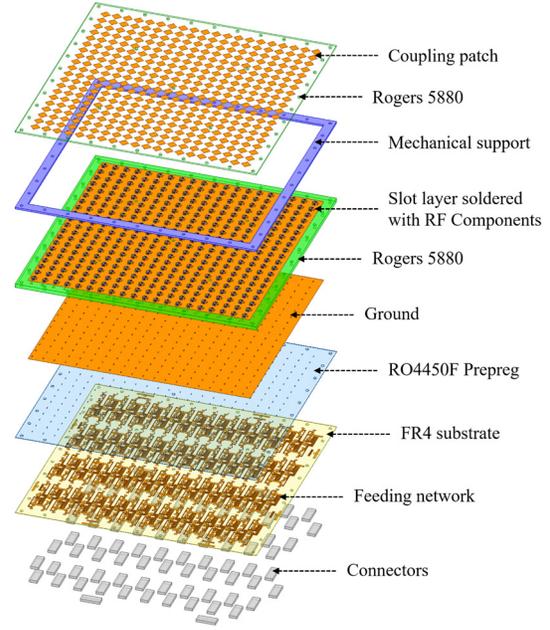

Fig. 9. 3D stacked array configuration of the designed 2-bit 5G mm-Wave RIS.

### B. Near Field Simulation

The full-wave simulation setup for the designed RIS array is shown in Fig. 10. In this simulation, the incident angle of the RIS system is assumed to be 30°, with the desired reflection beam at normal incidence (0°). A horn antenna (PE9851B/SF-2) is used as the excitation source for the RIS in the near field. The center operation frequency is set at 26 GHz. Overall, the near-field simulation for the entire RIS array aligns with [31] to ensure a fair comparison.

The phase distribution of the RIS can be theoretically calculated using the equation presented in [49]:

$$\varphi_{ij} = k \cdot |\,\vec{r}_{ij}^{\,e} - \vec{r}^{\,f}\,| - k \cdot (\vec{u}_0 \cdot \vec{r}_{ij}^{\,e}) + \Delta\varphi \qquad (2)$$

where $\varphi_{ij}$ is the ideal continuous phase distribution for the unit cell at the Cartesian coordinate system location $(i, j)$, with $i, j = 1...20$ for this design, $k$ represents the wavenumber at the free space, $r_{ij}^{\,e}$ refers to the element position vector, and $r^{\,f}$ dontes the excitation source position vector, $u_0$ indicates the unit vector of the desired reflected beam direction, and $\Delta\varphi$ is a flexible phase offset, which can adjust the reference phase level of the overall RIS distribution. Then, the 2-bit phase quantization can be expressed as:

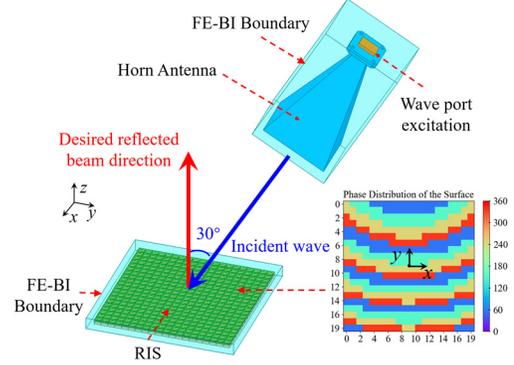

Fig. 10. The near-field full-wave simulation of the designed RIS.

$$\varphi_{ij} = \begin{cases} \varphi_1, \varphi_m = |\,\varphi_{ij} - \varphi_1\,| \\ \varphi_2, \varphi_m = |\,\varphi_{ij} - \varphi_2\,| \\ \varphi_3, \varphi_m = |\,\varphi_{ij} - \varphi_3\,| \\ \varphi_4, \varphi_m = |\,\varphi_{ij} - \varphi_4\,| \end{cases} \qquad (3)$$

$$\varphi_m = \min_{n \in \{1,2,3,4\}} \{|\,\varphi_{ij} - \varphi_n\,|\}$$

In other words, the ideal phase value $\varphi_{ij}$ is quantized to the nearest value among the four discrete states $\varphi_{1,2,3,4}$, which are obtained from the full-wave simulation of the unit cell (Fig. 6(b)). Thus, the RIS phase distribution with 2-bit phase quantization can be constructed, as shown in the inset of Fig. 10. After finalizing the simulation setup, the full-wave RIS simulation is conducted, and the simulated 3D radiation pattern is shown in Fig. 11. The simulation results indicate that the RIS achieves a gain of 24.4 dBi under the specified conditions. The 2D cross-sections of the E-plane and H-plane patterns will be presented in the next section, along with a comparison to the measured results.

### C. Far Field Simulation

The far-field radar cross-section (RCS) simulation setup for the designed RIS is shown in Fig. 12 (a). The excitation is changed to a plane wave illumination, where the wave is incident from the normal direction (0°) and the desired reflection angle is 30°. The polarization of the incident wave is along the *x*-axis, and the wave propagates along the −*z*-axis. The entire RIS is enclosed within a radiation boundary.

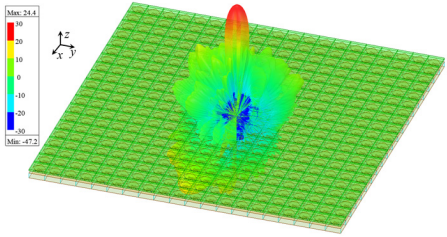

Fig. 11. Simulated 3D radiation pattern of the designed RIS.

For the sake of comparison, the far-field full-wave simulations of the RCS performance for the designed RIS with 1-bit phase quantization is shown in Fig. 12(b). From the simulation results, it can be observed that for 1-bit phase quantization, in addition to the desired main lobe at 30°, an obvious quantization lobe appears at -30°. In contrast, when the phase quantization is improved from 1-bit to 2-bit, the quantization lobe disappears in the far-field RCS response. The 2D H-plane (*yoz*-plane) RCS response for both 1-bit and 2-bit phase quantization are compared in Fig. 13. The results show that for 1-bit phase quantization, the far-field RCS response at the desired main lobe direction (30°) is 0.9 dBsm, while for 2-bit phase quantization, it increases to 5.9 dBsm. More importantly, the 1-bit quantization suffers from a quantization lobe level of 1.5 dBsm, which is at the same level as the main lobe. In comparison, this value decreases to -13.3 dBsm for 2-bit phase quantization.

### D. RIS Control Circuit Design

Finally, a versatile control circuit for RIS biasing applications with integral digital control and positive-negative voltage isolation functions is designed, and the schematic is shown in Fig. 14. Specifically, the digital control circuits include multiple shift registers (74HC595 chips) that multiplex the I/O of the microcontroller unit, making it suitable for large-scale RIS control circuits to reduce I/O usage. Additionally, a dual-polarity voltage isolation circuit is incorporated, using two optocouplers, as shown in Fig. 14(b). The outputs of the DC signal lines from the shift registers are connected to the inputs of the optocouplers. One optocoupler is responsible for driving the positive voltage bias to the PIN diode with a current-limiting resistor, while the other controls the negative voltage bias. This configuration allows independent control of the bias voltage for the two states of the PIN switches. Blue and red light-emitting diodes (LEDs) are connected in series with the inputs of the optocouplers to visually indicate the ON/OFF states of each PIN diode, facilitating code debugging. The ESP-WROOM-32 module is used to calculate the RIS pattern distribution and drive the overall control circuits. The complete diagram of the proposed RIS design with the control circuit biasing scheme is summarized in Fig. 15. Consequently, input commands can be coded on the PC side and transmitted via the real-time operating system (RTOS) of the embedded system, enabling real-time control of the RIS pattern distribution with low latency and deterministic behavior.

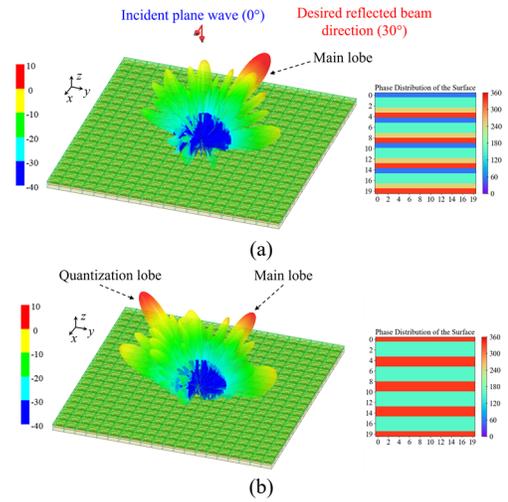

(a)

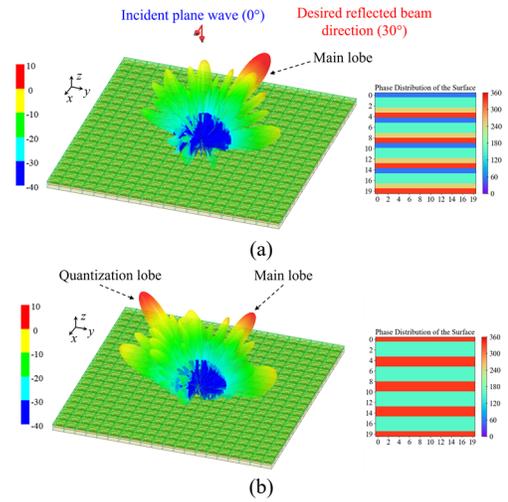

(b)

Fig. 12. Simulated 3D far-field RCS response of the designed RIS. (a) 2-bit phase quantization. (b) 1-bit phase quantization.

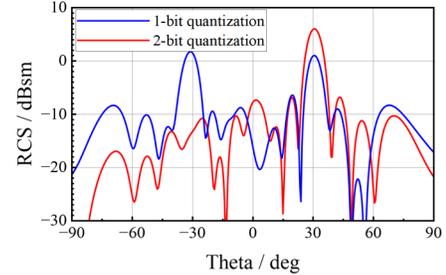

Fig. 13. Comparison of 2D H-plane far-field RCS response of the designed RIS with 1-bit and 2-bit phase quantization.

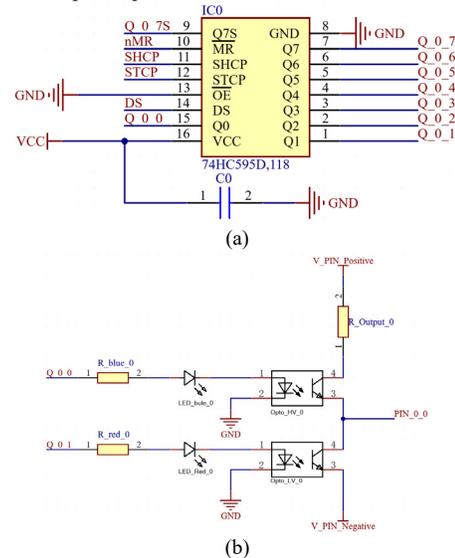

(a)

(b)

Fig. 14. The designed versatile control circuit for RIS biasing. (a) Digital control with shift registers. (b) Positive-negative voltage isolation circuit with optocouplers.

## III. FABRICATION AND MEASUREMENT

### A. RIS Fabricated Prototype

The fabricated 2-bit 5G mm-Wave RIS array board prototype is shown in Fig. 16. Specifically, the 1.575mm-thick Rogers 5880 substrate is laminated with an FR4 substrate, and PIN diodes and RF capacitors are soldered onto the top surface of the slot layer. The bottom layer contains the feeding network with flat flexible cable (FFC) connectors. The overall lamination board is depicted in Fig. 16(a). Another fabricated board consists of coupling patches on the 0.508mm-thick Rogers 5880 substrate, where the patches are etched on the bottom side. This board not only serves as the substrate for the patches but also provides protection (kind of packaging) for the rest of the RIS layers. The metallic patterns of the RIS boards are covered with a specially designed solder mask, ensuring high-quality component soldering. Additionally, the solder mask helps prevent any potential attachment issues during RIS assembly and minimizes the oxidation of the metallic patterns.

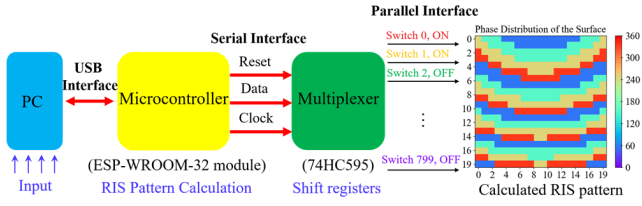

Fig. 15. Diagram of the overall RIS design with the control circuits.

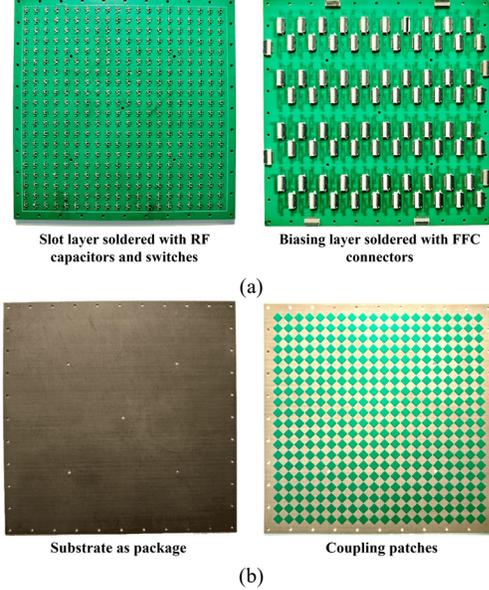

Fig. 16. Fabricated 2-bit 5G mm-Wave RIS prototype. (a) Slot and feeding network layers with soldered RF components and FFC connectors. (b) Coupling patch layer.

The fabricated prototype of the control circuit is shown in Fig. 17. The optocoupler and LED devices are routed on the top layer of the stacked PCB, which facilitates easy visualization of the RIS phase distribution. Meanwhile, the shift registers and FFC connectors are soldered onto the bottom layer of the stacked PCB. The same type of FFC connectors is used for both the RIS array board and the control circuit, which are interconnected through multiple FFCs. Similarly, the surface of the entire control circuit board is covered with a solder mask. It is important to note that the control circuit can be designed with multiple sub-boards and vertically assembled, further reducing fabrication costs and making the overall control circuit more compact. In this design, the control circuit is implemented on a single stacked board for clearer LED demonstration. Nevertheless, it does not impact the EM performance of RIS.

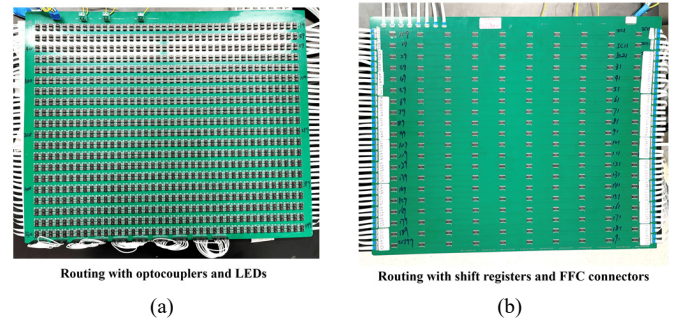

Fig. 17. Fabricated control circuit prototype. (a) Top view. (b) Bottom view.

Finally, the RIS array and control circuit boards are connected. Mechanical supports are printed using a Raise3D Pro2 printer with acrylonitrile butadiene styrene (ABS) filaments. The RIS array board is fixed vertically, while the control circuit board is placed horizontally to prevent electromagnetic wave interference.

### B. Measured Results and Comparison with Simulations for the Unit Cell

The practical phase performance and equivalent bit numbers are measured using the two-horn measurement setup [41], [45], [50]. Specifically, the RIS phase distribution is configured such that all unit cells are set to the same state, with the Tx and Rx horns symmetrically positioned along the RIS surface normal. For the 2-bit RIS design, there are four distinct phase states that need to be measured individually. This method is particularly effective at mm-Wave frequencies, as the small size of the unit cell can cause significant errors in waveguide measurement setups. The measurement setup for the designed 2-bit 5G mm-Wave RIS is shown in Fig. 18. The designed RIS is surrounded by thin mm-Wave absorbers to minimize interference reflections from the control circuits and the surrounding environment.

The measured individual phase curves for the four reconfigurable states and the measured equivalent bit numbers versus frequencies are shown in Fig. 19 (a) and (b), respectively. The measured results closely follow the simulation results across all four reconfigurable states. Furthermore, the measured results indicate that the equivalent 2-bit operation frequency band ($N_{bit} > 1.7$) spans from 24.1 to 27.7 GHz, with a relative bandwidth of 13.9%, which fully covers the 5G n258 band. Additionally, the measured effective 1-bit operating band extends from 20.0 to 28.9 GHz (36.4%). The slight deviations between the simulated and the measured results can be attributed to fabrication errors and RF components soldering tolerances, as well as the difference between the infinite periodic boundary conditions in simulation for the unit cell and the finite size of the RIS array in the measurement. Nonetheless, the wideband operation frequency band and steady phase transition curves have been successfully validated for the fabricated prototypes.

### C. Radiation Pattern Measurement in the Chamber

To further verify the consistency between simulation and measurement, the designed 2-bit 5G mm-Wave RIS was measured in an Orbit u-Lab mm-Wave anechoic chamber, as shown in Fig. 20. The measurement setup is consistent with the simulation configuration (Fig. 10), where the incident horn antenna is placed at a 30° angle, and the desired reflection beam is set to the normal direction (0°). The measured E-plane and H-plane radiation patterns are shown in Fig. 21 (a) and (b), respectively. The measured results are in good agreement with the simulation. The peak measured gain is 23.3 dBi, which is 1.1 dB lower than the simulation result, mainly due to losses from the practical prototype, including the RF capacitors, switches, and fabrication and assembly errors. It should be noted that for the H-plane radiation pattern, the gain drop around 20° − 30° is caused by incident horn blockage. Nevertheless, the anechoic chamber measurements further confirm the decent match between the simulation and measurement of the proposed 2-bit 5G mm-Wave RIS.

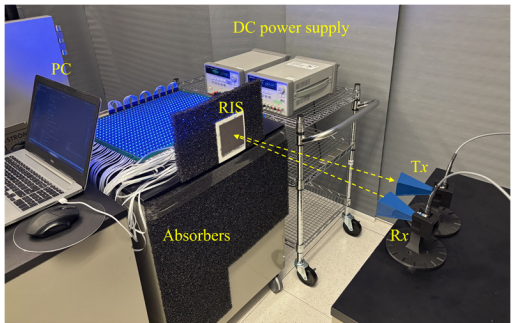

Fig. 18. Phase measurement setup for the designed 2-bit 5G mm-Wave RIS.

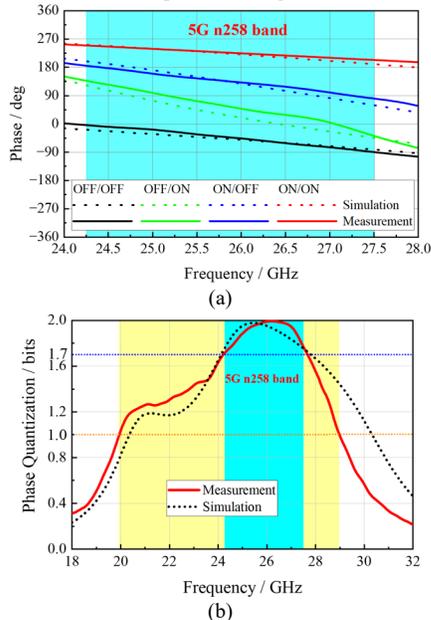

Fig. 19. Measured and simulated RIS reflection phase performance comparison. (a) Individual four different reconfigurable states. (b) Equivalent bits.

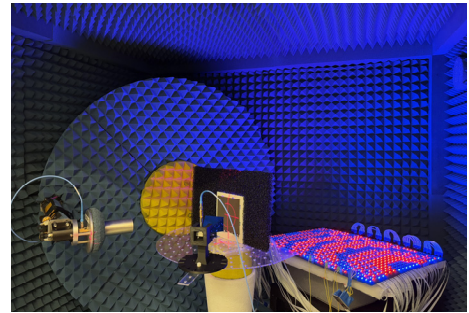

Fig. 20. Measurement setup in *u*-Lab mm-Wave anechoic chamber.

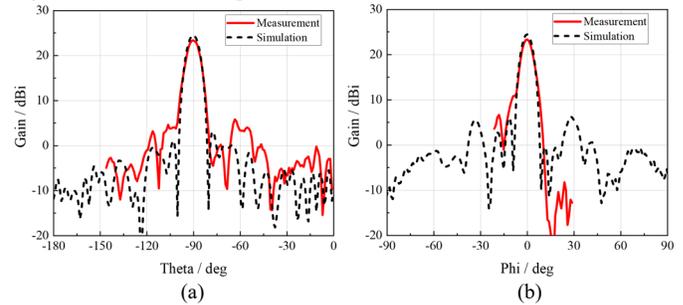

Fig. 21. Radiation patterns comparison at 26 GHz (a) E-plane. (b) H-plane.

### D. RIS Near Field Test

The near-field test setup is shown in Fig. 22. The Tx horn is positioned at an incident angle of 30° with a Tx-RIS distance of 25 cm. The desired reflection angle is 0° with a RIS-Rx distance of 35 cm. These two horns are connected to a PNA network analyzer E8363C, where the $S_{21}$ parameters indicate the received signal strength level. The supports for the horn antennas and RIS are 3D printed precisely to ensure they have the same height, meaning that the phase centers of the horns and the center of the RIS are located on the same horizontal plane. A laser-cut fixture with angle scale allows accurate control of the transmitting and receiving angles of the horns. Additionally, the RIS package is surrounded by absorbers to reduce interference from environmental objects. The RIS pattern distribution is visible from the blue and red LED arrays.

The comparison of the signal enhancement between the ON and OFF states is demonstrated in Fig. 23. At the center frequency of the designed RIS (26 GHz), a significant improvement from −38.3 dB to −9.9 dB (28.4 dB enhancement) is observed when the RIS pattern is turned ON, compared to when it is OFF. An average enhancement of 26.2 dB is achieved across the 5G mm-wave n258 frequency band, changed from −36.7 dB to −10.5 dB.

To validate the proper functioning of the designed RIS across the entire 5G n258 frequency band, detailed near-field measurements were conducted from 24.5 GHz to 27.5 GHz with a 1 GHz step. Specifically, The Rx antenna was moved to various observation angles from 0° to 80° in 5° steps to record the received signal level. The detailed comparison of the measured $S_{21}$ parameters in the ON and OFF conditions is shown in Fig. 24, with the 2-bit pattern distribution at various operating frequencies illustrated in the inset of each figure. It can be observed that the desired reflection beam levels remain consistent across the frequency range, with the measured $S_{21}$ parameters at -10.8, -9.8, -10.4, and -11.6 dB from 24.5 to 27.5 GHz, respectively. Thus, the peak gain variation is less than 1.8 dB. Additionally, the measured SLLs for the proposed RIS design are -15.5, -16.8, -14.6, and -14.7 dB for frequencies from 24.5 to 27.5 GHz. Therefore, an average SLL of -15.4 dB is achieved over the 5G n258 frequency range.

Furthermore, the beam scanning capability of the 2-bit RIS

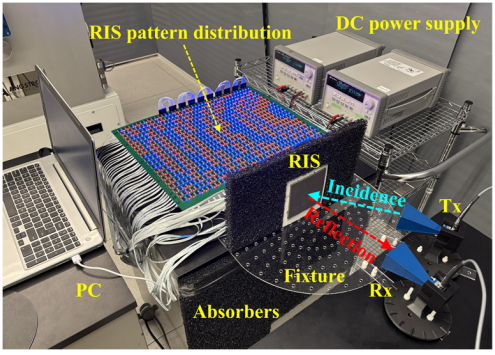

Fig. 22. Near-field characterization setup for the designed 2-bit 5G mm-Wave RIS.

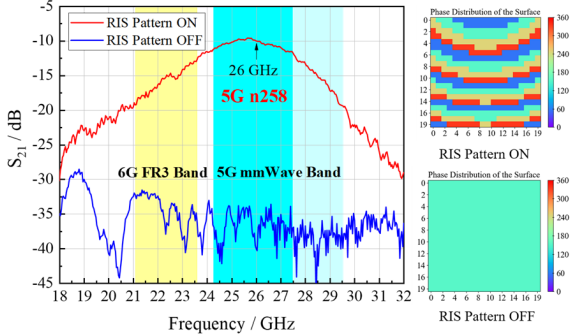

Fig. 23. Measured $S_{21}$ in ON/OFF condition with Tx-RIS distance of 25 cm and RIS-Rx distance of 35 cm.

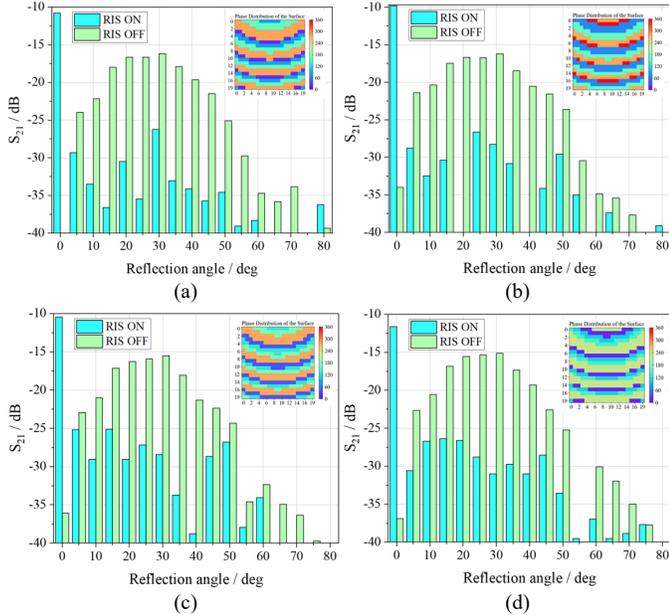

Fig. 24. Measured $S_{21}$ comparison of the RIS reflection when pattern on and off at different frequencies. (a) 24.5 GHz. (b) 25.5 GHz. (c) 26.5 GHz. (d) 27.5 GHz.

has been characterized and verified. In this setup, the operating frequency is fixed at the center frequency of 26 GHz, and the desired reflection beam is reconfigured from 0° to 50° by altering the RIS pattern distribution. The specific near-field beam scanning measurement results, comparing the RIS in the ON and OFF states, are shown in Fig. 25. The measured results reveal that a decent beam scanning capability is achieved from 0° to 50° in the RIS near field. The measured $S_{21}$ values at the desired reflection beam directions are -10.1, -10.0, -9.8, -9.7, -10.1, and -11.2 dB for angles from 0° to 50°, respectively. Meanwhile, the corresponding measured sidelobe levels (SLLs) are -15.4, -14.3, -15.2, -14.2, -14.8, and -13.0 dB. Therefore, both the wideband operational performance and the beam steering capability of the proposed RIS design have been validated, demonstrating higher gain enhancement and lower SLLs compared to the 1-bit counterpart.

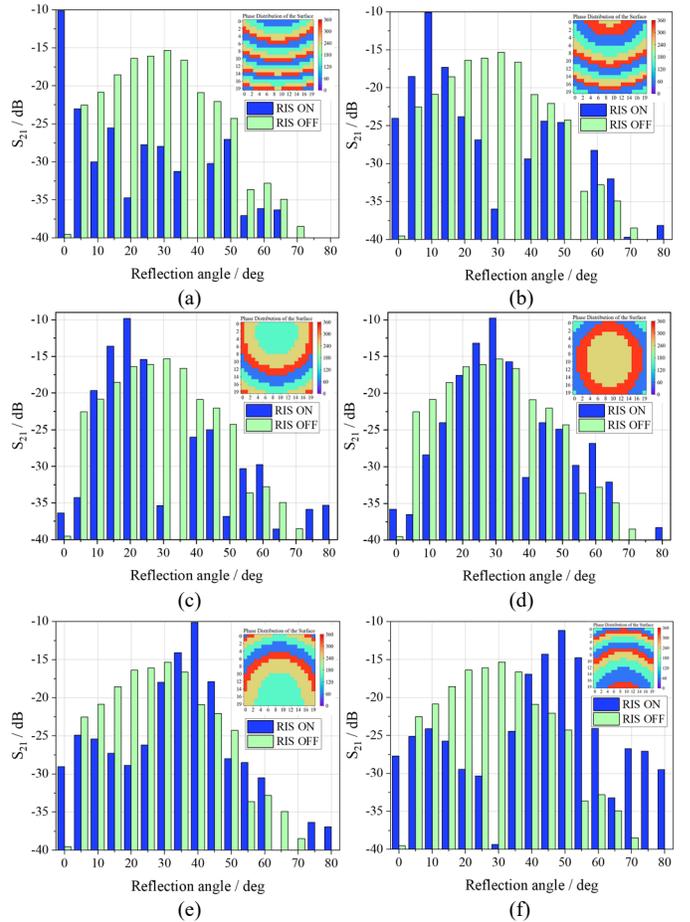

Fig. 25. Measured $S_{21}$ between the Tx and Rx horns when pattern ON and OFF at different reflected beam directions. (a) 0°. (b) 10°. (c) 20°. (d) 30°. (e) 40°. (f) 50°.

### E. RIS Far Field Test

To practically validate the quantization lobe suppression by the 2-bit mm-Wave RIS design, far-field measurements have been conducted. The detailed far-field measurement setup in a practical indoor environment is shown in Fig. 26. Consistent with the simulation setup in Fig. 12, the Tx horn antenna is placed at normal incidence angle and is connected to a PSG analog signal generator E8257D as the plane-wave excitation source. Meanwhile, the Rx horn is connected to a UXA signal analyzer N9040B to measure the received signal power. The Rx horn is moved to measure the received signal power level at both the main lobe direction (30°) and the quantization lobe direction (-30°). When the phase error across the entire RIS aperture is less than 22.5°, the illumination on the RIS can be considered as plane-wave. Therefore, the Tx should be placed at least at a distance of $2D^2/\lambda = 2.9$ m. Accordingly, the Tx-RIS and RIS-Rx distances are set to 3 m and 3.46 m in the experimental setup, respectively, ensuring proper plane-wave illumination.

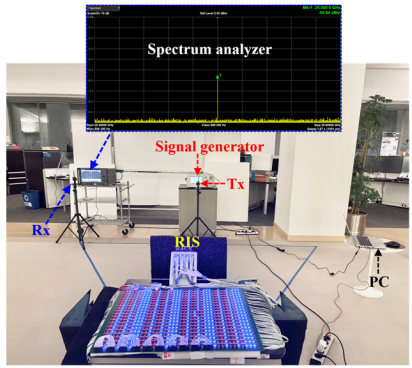

Fig. 26. Far-field characterization setup for the designed 2-bit 5G mm-Wave RIS.

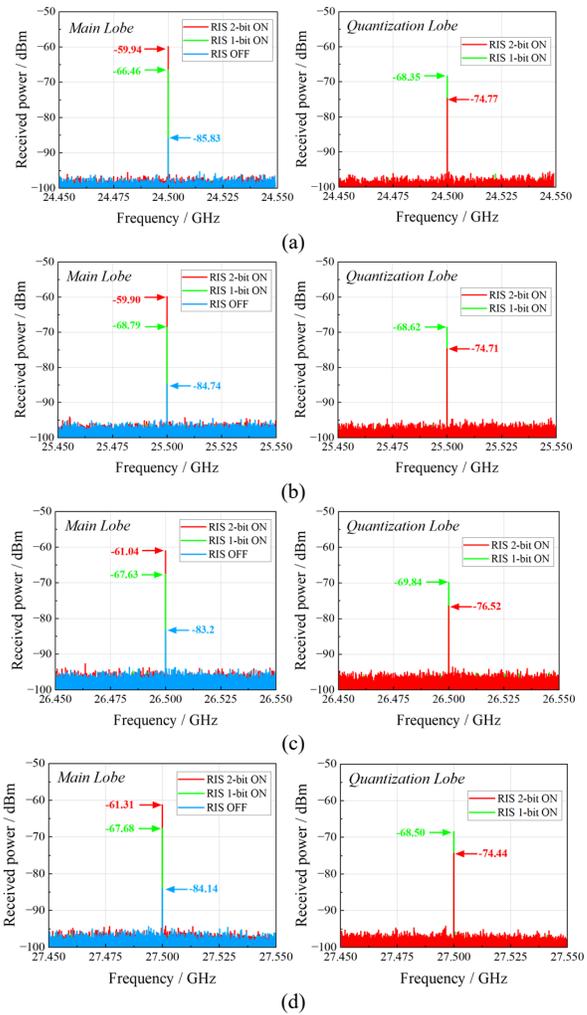

Fig. 27. Measured main lobe and quantization lobe levels with 1-bit and 2-bit phase quantization (a) 24.5 GHz. (b) 25.5 GHz. (c) 26.5 GHz. (d) 27.5 GHz.

The detailed measured received signal power levels for the main lobe and quantization lobes, from 24.5 GHz to 27.5 GHz with a 1 GHz frequency step, are shown in Fig. 27. On one hand, for the main lobes, the measured received power levels with 2-bit phase quantization are -59.9, -59.9, -61.0, and -61.3 dBm, which provide 25.9, 24.8, 22.2, and 22.8 dB power enhancement compared with the RIS pattern OFF state. Therefore, an average power enhancement of 23.9 dB is achieved in the RIS far field. Additionally, 2-bit phase quantization provides approximately 7 dB more power enhancement over the frequency band compared to 1-bit phase quantization. On the other hand, the QLLs are significantly suppressed with 2-bit phase quantization, with values of -14.8, -14.8, -15.5, and -13.1 dB relative to the main lobe within the frequency range of 24.5 GHz to 27.5 GHz, and an average QLL of -14.6 dB is obtained. In contrast, the average QLL for 1-bit phase quantization is -1.3 dB. Thus, the 2-bit phase quantization results in a measured 13.3 dB QLL suppression compared to 1-bit quantization, which is matched with the far-field RCS suppression observed in the simulation (Fig. 13).

### F. Comparison and Discussion

Finally, a comparison table is provided to demonstrate the performance of the proposed design and previously reported RIS designs, as summarized in Table II. To date, published practical RIS prototypes have primarily focused on low frequencies around or below 10 GHz [5], [13 – 25]. Only a few RIS hardware designs have targeted the 5G mm-Wave frequency bands. [26] designed a 5G mm-Wave RIS with significant gain enhancement, but it suffers from a narrow bandwidth of 7.3%. [31] reported a wideband 5G mm-Wave RIS design, but the 1-bit phase quantization resulted in a high sidelobe level of -7.8 dB in the near field and quantization lobes in the far field. In this work, we have designed, fabricated, and measured a 2-bit 5G mm-Wave RIS that demonstrates significantly better performance compared with [31], achieved by adding an additional switch to each unit cell. In summary, the proposed 2-bit RIS design offers a gain enhancement of 3.7 dB and 5 dB in the RIS near field and far field, respectively, with the same 20 × 20 RIS array configuration. Additionally,

### TABLE II
PERFORMANCE COMPARISON OF THE PROPOSED DESIGN WITH PREVIOUS RIS

| | Ref | [21] | [5] | [26] | [31] | **This work** |
|---|---|---|---|---|---|---|
| | Freq. (GHz) | 3.15 | 2.3 | 28.5 | 27.5 | **26** |
| | Phase quantization | 3-bit | 2-bit | 1-bit | 1-bit | **2-bit** |
| | Switch type, Switch numbers | Varactor, 1 | PIN, 5 | PIN, 1 | PIN, 1 | **PIN, 2** |
| | Size ($\lambda \times \lambda$) | 3×3 | 6.1×6.1 | 9.5×9.5 | 7.1×7.1 | **8.0×8.0** |
| | Measured 1-bit BW (%) | NA | NA | 7.3 | 26.9 | **36.4** |
| | Measured 2-bit BW (%) | 14 | 15.2 | 0 | 0 | **13.9** |
| Near field (dB) | Enhancement | NA | 21.9 | ~30 | 24.7 | **28.4** |
| | SLL | -4 | -12 | NA | -7.8 | **-15.4** |
| Far field (dB) | Enhancement | NA | NA | ~25 | 18.9 | **23.9** |
| | QLL | NA | NA | 0 | -1.3 | **-14.6** |
| | Power consumption (W) | NA | NA | 8 | 1 | **2** |

the designed 2-bit RIS achieves a 7.6 dB lower sidelobe level (SLL) as well as a 13.3 dB suppression of the far-field quantization lobe. The effective 1-bit bandwidth (BW) is improved to 36.4%, and the 2-bit BW reaches 13.9%, fully covering the 5G n258 frequency band. Therefore, the proposed RIS design is a promising candidate for 5G mm-Wave RIS-aided communication systems with low interference requirements.

## IV. Conclusion

In this work, a wideband 5G mm-Wave RIS with 2-bit phase quantization for the 5G n258 band is proposed. The design demonstrates a measured effective 13.9% 2-bit bandwidth (BW) and a 36.4% 1-bit BW. Furthermore, it shows significant overall performance improvements over the 1-bit counterpart, both in the RIS near and far fields. It is noteworthy that the proposed RIS solution utilizes only 2 PIN switches in each unit, which does not significantly increase design complexity. Therefore, the proposed 5G mm-Wave RIS hardware solution is a promising candidate for practical advanced mm-Wave RIS-aided communication systems.

## References

[1] W. Hong et al., "Multibeam Antenna Technologies for 5G Wireless Communications," *IEEE Trans. Antennas Propag.*, vol. 65, no. 12, pp. 6231-6249, Dec. 2017.

[2] T. S. Rappaport, Y. Xing, G. R. MacCartney, A. F. Molisch, E. Mellios, and J. Zhang, "Overview of millimeter wave communications for fifth-generation (5G) wireless networks—With a focus on propagation models," *IEEE Trans. Antennas Propag.*, vol. 65, no. 12, pp. 6213–6230, Dec. 2017.

[3] S. Hur et al., "Proposal on millimeter-wave channel modeling for 5G cellular system," *IEEE J. Sel. Topics Signal Process.*, vol. 10, no. 3, pp. 454–469, Apr. 2016.

[4] I. F. Akyildiz, C. Han, and S. Nie, "Combating the distance problem in the millimeter wave and terahertz frequency bands," *IEEE Commun. Mag.*, vol. 56, no. 6, pp. 102–108, Jun. 2018.

[5] L. Dai et al., "Reconfigurable intelligent surface-based wireless communications: Antenna design, prototyping, and experimental results," *IEEE Access*, vol. 8, pp. 45913–45923, 2020.

[6] X. Cao, Q. Chen, T. Tanaka, M. Kozai, and H. Minami, "A 1-bit time-modulated reflectarray for reconfigurable-intelligent-surface applications," *IEEE Trans. Antennas Propag.*, vol. 71, no. 3, pp. 2396–2408, Mar. 2023.

[7] J. Rao et al., "A Shared-Aperture Dual-Band Sub-6 GHz and mmWave Reconfigurable Intelligent Surface With Independent Operation," *IEEE Trans. Microw. Theory Techn.*, Dec. 2024. doi: 10.1109/TMTT.2024.3506218.

[8] Y. Liu et al., "Reconfigurable intelligent surfaces: Principles and opportunities," *IEEE Commun. Surveys Tuts.*, vol. 23, no. 3, pp. 1546–1577, 2nd Quart., 2021.

[9] P. Zheng, X. Liu, and T. Y. Al-Naffouri, "LEO- or RIS-empowered user tracking: A Riemannian manifold approach," *IEEE J. Sel. Areas Commun.*, vol. 42, no. 12, pp. 3445–3461, 2024.

[10] X. Mu, Y. Liu, L. Guo, J. Lin and R. Schober, "Simultaneously Transmitting and Reflecting (STAR) RIS Aided Wireless Communications," *IEEE Trans. Wireless Commun.*, vol. 21, no. 5, pp. 3083-3098, May 2022.

[11] H. Li, S. Shen and B. Clerckx, "Beyond Diagonal Reconfigurable Intelligent Surfaces: From Transmitting and Reflecting Modes to Single-, Group-, and Fully-Connected Architectures," *IEEE Trans. Wireless Commun.*, vol. 22, no. 4, pp. 2311-2324, April 2023.

[12] H. Chen, P. Zheng, M. F. Keskin, T. Al-Naffouri, and H. Wymeersch, "Multi-RIS-Enabled 3D sidelink positioning," *IEEE Trans. Wireless Commun.*, vol. 23, no. 8, pp. 8700–8716, Aug. 2024.

[13] D. Inserra et al., "Dual-Orthogonal Polarization Amplifying Reconfigurable Intelligent Surface With Reflection Amplifier Based on Passive Circulator," *IEEE Trans. Microw. Theory Techn.*, vol. 72, no. 7, pp. 4383-4394, July 2024.

[14] R. Song, H. Yin, Z. Wang, T. Yang and X. Ren, "Modeling, Design, and Verification of an Active Transmissive RIS," *IEEE Trans. Antennas Propag.*, vol. 72, no. 12, pp. 9239-9250, Dec. 2024.

[15] J. Tian, H. Yang, T. Li, Z. Zhang, J. Han and X. Cao, "Realization and Analysis of Low-Loss Reconfigurable Quasi-Periodic Coding Metasurfaces for Low-Cost Single-Beam Scanning," *IEEE Trans. Microw. Theory Techn.*, vol. 72, no. 9, pp. 5071-5081, Sept. 2024.

[16] Z. Zhang et al., "A Novel Design Approach Using Zero-Pole-Based Multiport Model for Reconfigurable Intelligent Surfaces," *IEEE Trans. Antennas Propag.*, vol. 72, no. 11, pp. 8564-8574, Nov. 2024.

[17] Q. Hu, H. Yang, X. Zeng, Y. Rao and X. Y. Zhang, "Methodology and Design of Absorptive Filtering Reconfigurable Intelligent Surfaces," *IEEE Trans. Antennas Propag.*, vol. 72, no. 6, pp. 5301-5306, June 2024.

[18] J. Rao, Y. Zhang, S. Tang, Z. Li, C.-Y. Chiu, and R. Murch, "An active reconfigurable intelligent surface utilizing phase-reconfigurable reflection amplifiers," *IEEE Trans. Microw. Theory Techn.*, vol. 71, no. 7, pp. 3189–3202, Jul. 2023.

[19] A. Sayanskiy et al., "A 2D-programmable and scalable reconfigurable intelligent surface remotely controlled via digital infrared code," *IEEE Trans. Antennas Propag.*, vol. 71, no. 1, pp. 570–580, Jan. 2023.

[20] L. Wu et al., "A wideband amplifying reconfigurable intelligent surface," *IEEE Trans. Antennas Propag.*, vol. 70, no. 11, pp. 10623–10631, Nov. 2022.

[21] J. C. Liang et al., "An angle-insensitive 3-bit reconfigurable intelligent surface," *IEEE Trans. Antennas Propag.*, vol. 70, no. 10, pp. 8798–8808, Oct. 2022.

[22] X. Pei et al., "RIS-aided wireless communications: Prototyping, adaptive beamforming, and indoor/outdoor field trials," *IEEE Trans. Commun.*, vol. 69, no. 12, pp. 8627–8640, Dec. 2021.

[23] Q. Hu, H. Yang, Y. Rao and X. Y. Zhang, "Dual-Band Aperture-Shared Filtering Reconfigurable Intelligent Surface," *IEEE Trans. Antennas Propag.*, Dec. 2024. doi: 10.1109/TAP.2024.3520797.

[24] Y. Zhao et al., "2-Bit RIS Prototyping Enhancing Rapid-Response Space-Time Wavefront Manipulation for Wireless Communication: Experimental Studies," *IEEE Open J. Commun. Soc.*, vol. 5, pp. 4885-4901, 2024, doi: 10.1109/OJCOMS.2024.3439558.

[25] Z. Zhang et al., "Macromodeling of reconfigurable intelligent surface based on microwave network theory," *IEEE Trans. Antennas Propag.*, vol. 70, no. 10, pp. 8707–8717, Oct. 2022.

[26] J.-B. Gros, V. Popov, M. A. Odit, V. Lenets, and G. Lerosey, "A reconfigurable intelligent surface at mmWave based on a binary phase tunable metasurface," *IEEE Open J. Commun. Soc.*, vol. 2, pp. 1055–1064, 2021.

[27] A. S. Shekhawat, B. G. Kashyap, R. W. R. Torres, F. Shan and G. C. Trichopoulos, "A Millimeter-Wave Single-Bit Reconfigurable Intelligent Surface with High-Resolution Beam-Steering and Suppressed Quantization Lobe," *IEEE Open J. Antennas Propag.*, doi: 10.1109/OJAP.2024.3506453.

[28] W. Tang et al., "Path Loss Modeling and Measurements for Reconfigurable Intelligent Surfaces in the Millimeter-Wave Frequency Band," *IEEE Trans. Commun.*, vol. 70, no. 9, pp. 6259-6276, Sept. 2022.

[29] H. Radpour, M. Hofer, L. W. Mayer, A. Hofmann, M. Schiefer, and T. Zemen, "Active reconfigurable intelligent surface for the millimeterwave frequency band: Design and measurement results," in *Proc. IEEE Wireless Commun. Netw. Conf. (WCNC)*, Apr. 2024, pp. 1–6.

[30] A. S. Shekhawat, B. G. Kashyap, B. Tjahjadi, and G. C. Trichopoulos, "Beamforming characterization of a mmWave single-bit reflective metasurface," in *Proc. IEEE Int. Symp. Antennas Propag. USNC-URSI Radio Sci. Meeting (AP-S/URSI)*, Jul. 2022, pp. 1608–1609.

[31] R. Wang, Y. Yang, B. Makki and A. Shamim, "A Wideband Reconfigurable Intelligent Surface for 5G Millimeter-Wave Applications," *IEEE Trans. Antennas Propag.*, vol. 72, no. 3, pp. 2399-2410, Mar. 2024.

[32] H. Yang et al., "A 1-bit 10 × 10 reconfigurable reflectarray antenna: Design, optimization, and experiment," *IEEE Trans. Antennas Propag.*, vol. 64, no. 6, pp. 2246–2254, Jun. 2016.

[33] X. Pan, F. Yang, S. Xu, and M. Li, "A 10 240-element reconfigurable reflectarray with fast steerable monopulse patterns," *IEEE Trans. Antennas Propag.*, vol. 69, no. 1, pp. 173–181, Jan. 2021.

[34] F. Wu, R. Lu, J. Wang, Z. H. Jiang, W. Hong, and K.-M. Luk, "A circularly polarized 1 bit electronically reconfigurable reflectarray based on electromagnetic element rotation," *IEEE Trans. Antennas Propag.*, vol. 69, no. 9, pp. 5585–5595, Sep. 2021.

[35] H. Luyen, Z. Zhang, J. H. Booske, and N. Behdad, "Wideband, beamsteerable reflectarray antennas exploiting electronically

reconfigurable polarization-rotating phase shifters," *IEEE Trans. Antennas Propag.*, vol. 70, no. 6, pp. 4414–4425, Jun. 2022.

[36] K. Duan et al., "A 2-bit Dual-Channel Reconfigurable Reflectarray with Beam Scanning Capability," *IEEE Trans. Antennas Propag.*, Nov, 2024.

[37] H. Yu et al., "Quad-Polarization Reconfigurable Reflectarray With Independent Beam-Scanning and Polarization Switching Capabilities," *IEEE Trans. Antennas Propag.*, vol. 71, no. 9, pp. 7285-7298, Sept. 2023.

[38] L. Zhu et al., "Dual Linearly Polarized 2-bit Programmable Metasurface With High Cross-Polarization Discrimination," *IEEE Trans. Antennas Propag.*, vol. 72, no. 2, pp. 1510-1520, Feb. 2024.

[39] A. Nasser et al., "Online DRL-based Beam Selection for RIS-Aided Physical Layer Security: An Experimental Study" in *Proc. IEEE Global Commun. Conf.*, Dec. 2024.

[40] M. Frank, F. Lurz, R. Weigel and A. Koelpin, "Electronically Reconfigurable 6 × 6 Element Transmitarray at K-Band Based on Unit Cells With Continuous Phase Range," *IEEE Antennas Wireless Propag. Lett.*,vol. 18, no. 4, pp. 796-800, Apr. 2019.

[41] C. Liu, F. Yang, S. Xu and M. Li, "An E-Band Reconfigurable Reflectarray Antenna Using p-i-n Diodes for Millimeter-Wave Communications," *IEEE Trans. Antennas Propag.*, vol. 71, no. 8, pp. 6924-6929, Aug. 2023.

[42] P. Zheng, R. Wang, A. Shamim and T. Y. Al-Naffouri, "Mutual Coupling in RIS-Aided Communication: Model Training and Experimental Validation," *IEEE Trans. Wireless Commun.*, vol. 23, no. 11, pp. 17174-17188, Nov. 2024.

[43] H. Yang et al., "A study of phase quantization effects for reconfigurable reflectarray antennas," *IEEE Antennas Wireless Propag. Lett.*, vol. 16, pp. 302–305, 2017.

[44] H. Luyen, J. H. Booske and N. Behdad, "2-Bit Phase Quantization Using Mixed Polarization-Rotation/Non-Polarization- Rotation Reflection Modes for Beam-Steerable Reflectarrays," *IEEE Trans. Antennas Propag.*, vol. 68, no. 12, pp. 7937-7946, Dec. 2020.

[45] E. Wang et al., "A 1296-Cell Reconfigurable Reflect-Array Antenna With 2-bit Phase Resolution for Ka-Band Applications," *IEEE Trans. Antennas Propag.*, vol. 72, no. 4, pp. 3425-3437, April 2024, doi: 10.1109/TAP.2024.3368220.

[46] J. Tang, M. Cui, S. Xu, L. Dai, F. Yang and M. Li, "Transmissive RIS for B5G Communications: Design, Prototyping, and Experimental Demonstrations," *IEEE Trans. Commun.*, vol. 71, no. 11, pp. 6605-6615, Nov. 2023.

[47] Y. Yin et al., "Design of a 2-bit Dual-Polarized Reconfigurable Reflectarray With High Aperture Efficiency," *IEEE Trans. Antennas Propag.*, vol. 72, no. 1, pp. 542-552, Jan. 2024, doi: 10.1109/TAP.2023.3326951.

[48] R. Pereira, R. Gillard, R. Sauleau, P. Potier, T. Dousset, and X. Delestre, "Dual linearly-polarized unit-cells with nearly 2-bit resolution for reflectarray applications in X-band," *IEEE Trans. Antennas Propag.*, vol. 60, no. 12, pp. 6042–6048, Dec. 2012.

[49] P. Nayeri, F. Yang, and A. Z. Elsherbeni, *Reflectarray Antennas: Theory, Designs, and Applications*. Hoboken, NJ, USA: Wiley, 2018.

[50] Z. H. Jiang, L. Kang, T. Yue, W. Hong, and D. H. Werner, "Wideband transmit arrays based on anisotropic impedance surfaces for circularly polarized single-feed multibeam generation in the Q-band," *IEEE Trans. Antennas Propag.*, vol. 68, no. 1, pp. 217–229, Jan. 2020.